\documentclass[%
 reprint,
 amsmath,amssymb,
 aps,
prb,
floatfix,
]{revtex4-2}

\usepackage{graphicx}
\usepackage{dcolumn}
\usepackage{subfigure}
\usepackage{bm}
\usepackage{hyperref}
\usepackage[T2A,T1]{fontenc}
\usepackage[utf8]{inputenc}
\usepackage{float}

\usepackage{amsmath,amssymb}
\usepackage{lmodern}
\usepackage{mathtools}
\usepackage{bm}
\usepackage{physics}
\usepackage{xcolor}
\usepackage{graphicx}
\usepackage{calc}

\usepackage{hyperref}
\hypersetup{
    colorlinks=true,
    linkcolor=red,
    citecolor=red,
    filecolor=magenta,      
    urlcolor=blue,
    }

\begin{document}

\title{Resonant state expansion for acoustic resonators. \\ Part I. Eigenvalue problem}

\author{Egor~Domoratskii$^{1,2}$}

\author{Vladimir~Igoshin$^{2}$}

\author{Nikolay Solodovchenko$^{1,2,3}$}

\author{Mingzhao~Song$^{1}$}

\author{Yong~Li$^{2,4}$}
\email{yongli@tongji.edu.cn}

\author{Mihail~Petrov$^{2}$}
\email{m.petrov@metalab.ifmo.ru}

\author{Andrey Bogdanov$^{1,3}$}
\email{a.bogdanov@hrbeu.edu.cn}

\affiliation{$^{1}$ Qingdao Innovation and Development Center, Harbin Engineering University, Qingdao 266000, China}

\affiliation{$^{2}$ School of Physics and Engineering, ITMO University, St. Petersburg 197101, Russia}

\affiliation{$^{3}$ Ioffe Institute, St. Petersburg 194021, Russia}

\affiliation{$^{4}$ Institute of Acoustics, School of Physics Science and Engineering, Tongji University, Shanghai 200092, China}

\begin{abstract}
Resonant-state expansion (RSE) is a powerful modal framework for the perturbative analysis of open resonant systems, providing direct access to complex eigenfrequencies and eigenmodes. While RSE is well developed in electromagnetism, a comparably systematic formulation for acoustics remains less established. Here, we develop a general Green-function-based formalism for acoustic RSE and illustrate it for a class of two-dimensional acoustic resonators. Using the resonant states of an analytically solvable cylindrical reference system as a basis, we derive explicit perturbation matrix elements for uniform, radial, and sectoral variations of density and compressibility, representing homogeneous tuning, graded profiles, and symmetry-induced modal coupling. The resulting complex eigenfrequencies and eigenmodes are validated against exact analytical solutions and finite-element simulations, showing excellent quantitative agreement. The framework provides a systematic and physically transparent approach for analyzing perturbed open acoustic resonators and establishes a basis for resonant-state methods in acoustic metamaterials and non-Hermitian acoustics.
\end{abstract}

\maketitle

\section{Introduction}

Acoustic metamaterials and phononic crystals provide versatile routes for controlling sound through structures whose response is governed by resonances and engineered effective material properties~\cite{lu2009phononic}. Subwavelength acoustic control has been demonstrated using both locally resonant structures and space-coiling geometries, which enable strong resonance effects and compact manipulation of sound propagation~\cite{liu2000locally,liang2012extreme}. These ideas have since developed into a broad framework for subwavelength sound manipulation using resonators, metamaterials, and metastructures~\cite{cummer2016controlling}, including resonant and metamaterial approaches to sound absorption~\cite{huang2023sound,yang2017sound,gao2022acoustic,wang2024stacked}. At the level of individual building blocks, the eigenmodes of compact resonators, together with their symmetry and multipolar content, determine much of the response of larger resonant structures~\cite{tsimokha2022acoustic,gladyshev2020symmetry}. For open resonators coupled to propagating waves in the surrounding medium, the eigenmodes differ from the normal modes of a closed Hermitian cavity. Outgoing-wave boundary conditions account for radiation into the environment and lead to resonances with complex eigenfrequencies~\cite{huang2024acoustic}. This open-system spectral structure is particularly important in modern acoustic resonators supporting high-$Q$ states, bound states in the continuum (BICs), quasi-BICs, exceptional points, and directional scattering~\cite{lyapina2015bound, jia2023bound, deriy2022bound,achilleos2017nonhermitian,igoshin2024exceptional,krasikova2024acoustic,zhu2018simultaneous,timankova2026experimental,PhysRevApplied.22.064041}. Closely related BIC physics has been extensively developed in open photonic structures~\cite{koshelev2021bound}, providing useful conceptual parallels for resonant acoustic systems. A modal framework intended for such systems must therefore treat the outgoing boundary condition and complex resonant spectrum together with a normalization appropriate to open modes, rather than importing the ordinary energy normalization and Hermitian orthogonality of closed cavities.

A systematic framework of this type is well developed in electromagnetism through the resonant-state expansion (RSE). The underlying approach was introduced as a Brillouin-Wigner perturbation theory for open electromagnetic systems, in which a modified normalization of resonant states permits a spectral representation of the Green's function and converts the perturbed problem into a matrix eigenvalue problem~\cite{muljarov2011brillouin}. The RSE was subsequently developed for two-dimensional open systems, where the Green's function of a dielectric cylinder contains a branch cut that must be incorporated into the spectral basis~\cite{doost2013resonant}, for three-dimensional open optical resonators~\cite{doost2014resonant}, and for planar photonic-crystal structures~\cite{neale2020resonant}. Further extensions include frequency-dispersive materials, magnetic, chiral, and bi-anisotropic media~\cite{muljarov2018resonant, muljarov2016resonant} with related applications to chiral and bianisotropic resonators~\cite{both2022nanophotonic,shakirova2025molecular,poleva2023multipolar}, as well as the calculation of scattering observables from resonant-state Green-function expansions~\cite{lobanov2018resonant}. Static contributions require particular care in open-system spectral representations. A complete set of static modes for three-dimensional optical RSE was formulated in Ref.~\cite{lobanov2019resonant}, while subsequent work developed a full electromagnetic Green's dyadic formulation that allows static modes to be eliminated from the RSE~\cite{muljarov2020full}. More broadly, the theory of resonant states (RSs) and quasinormal modes (QNMs) in open electromagnetic systems has been reviewed from the perspectives of light-resonance interaction, resonant-state expansions, and the normalization, orthogonality, and completeness of QNMs~\cite{lalanne2018light,both2022resonant,sauvan2022normalization}. These works emphasize that open resonances require generalized normalization and that a spectral representation may contain continuum, branch-cut, or additional static contributions depending on the analytic structure of the problem. Thus, the appropriate mathematical object is a generalized resonant-state expansion rather than an ordinary orthogonal normal-mode expansion.

Several approaches have been developed to calculate acoustic eigenfrequencies and eigenmodes in specific geometries, including spherical cavities with eccentric inner inclusions, spheroidal cavities containing penetrable spheres, and weakly deformed quasispherical resonators~\cite{roumeliotis1992acoustic,kokkorakis1999acoustic,mehl2007acoustic}. Such methods can provide accurate analytical or semi-analytical descriptions for the geometries for which they are derived, but they do not by themselves constitute a general perturbative framework for reusing the resonant spectrum of an open reference system under arbitrary material or geometric perturbations. Related open-mode approaches have increasingly been transferred to mechanical and elastic wave systems. QNM descriptions have been developed for dissipative optomechanical cavities, radiating resonators in open phononic systems, and thin elastic plates~\cite{elsayed2020quasinormal,laude2023quasinormal,vial2024quasinormal}, building on the QNM formalism developed for open photonic and plasmonic resonators~\cite{sauvan2013theory,lalanne2018light}. Resonant-state ideas have also already been employed in acoustics: in particular, the exceptional-point problem of a single open acoustic resonator was analyzed using a reduced two-mode RSE-based perturbative model~\cite{igoshin2024exceptional}.

Transferring the full RSE formalism from electromagnetism to acoustics is nevertheless nontrivial. The natural first-order acoustic state consists of the pressure $p$ and the vector particle velocity $\mathbf{v}$, while the constitutive operator is governed by density and compressibility rather than permittivity and permeability. Moreover, the acoustic Green's function and the outgoing-wave problem have their own spectral and normalization structure, so the corresponding resonant-state normalization and perturbation matrix elements must be derived specifically for the acoustic system. This distinction is important for modern resonant acoustic structures such as gradient-index labyrinthine systems, multi-sector resonators, slow-sound media, and compact metamaterial elements, where relatively small material or structural perturbations can produce substantial shifts of resonant frequencies and $Q$ factors and strong modal hybridization. Although such effects can be studied numerically or with reduced coupled-mode descriptions, a systematic full-wave perturbative framework based on the resonant states of a single open reference resonator remains highly desirable.

In this work (Part I), we formulate such a Green-function-based RSE for open acoustic resonators and demonstrate it using an analytically solvable two-dimensional cylindrical resonator as the reference system. Starting from the acoustic equations in a first-order pressure--velocity representation, we derive the resonant Green's function and the normalization of the acoustic resonant states and obtain the matrix eigenvalue problem governing perturbations of the density and compressibility. We explicitly account for the branch-cut contribution required by the two-dimensional open problem. The resulting framework is then applied to uniform material perturbations, radially varying perturbations, and sectoral perturbations that break cylindrical symmetry and couple different azimuthal orders. These cases probe, respectively, homogeneous constitutive tuning, radially graded material profiles, and symmetry-controlled mode coupling and hybridization. Comparison with independent finite-element eigenfrequency calculations is used to assess the accuracy of the expansion for the perturbations considered. To our knowledge, the present work provides a systematic Green-function-based formulation of the resonant-state expansion for open acoustic resonators, including explicit acoustic resonant-state normalization and perturbation matrices for density and compressibility. The formulation provides a starting point for extending resonant-state techniques to more complex open acoustic resonators and, in future work, to scattering calculations and more general geometries.

\section{Resonant-State Expansion for Open Acoustic Resonators}
\label{sec:resonant_state_expansion}
\subsection{Acoustic resonant states and Green's function}

In general, the linear acoustic equations for the media with density $\rho_\omega(\mathbf{x})$ and $\beta_\omega(\mathbf{x})$ compressibility have the following form:
\begin{equation}
    \begin{cases}
        i \omega \beta_\omega(\mathbf{x}) p(\mathbf{x}) = \nabla \cdot \mathbf{v}(\mathbf{x}), \\
        i \omega \rho_\omega(\mathbf{x}) \mathbf{v}(\mathbf{x}) = \nabla p(\mathbf{x}) + \mathbf{f}(\mathbf{x}),
    \end{cases}
    \label{eq:eq_system_complex}
\end{equation}
where $p(\mathbf{x})$ -- pressure, $\mathbf{v}(\mathbf{x})$ -- velocity,  and  $\mathbf{f}(\mathbf{x})$ is the force volume density~\cite{landau1987fluid, toftul2019acoustic}. Time-dependent parts of pressure and velocity are given by $\exp(-i \omega t)$ with the complex frequency $\omega$. The material parameters are connected with each other by the sound velocity $c_\omega(\mathbf{x}) = [\rho_\omega(\mathbf{x}) \beta_\omega(\mathbf{x})]^{-1/2}$. Here and further, the subscript $\omega$ denotes frequency dependence. 

Let us write Eqs.~\eqref{eq:eq_system_complex} in the following inhomogeneous matrix form
\begin{equation}
    \hat{\mathbb{D}}(\mathbf{x}) \vec{\mathbb{F}}(\mathbf{x}) = \omega \hat{\mathbb{P}}_\omega(\mathbf{x}) \vec{\mathbb{F}}(\mathbf{x}) - \vec{\mathbb{J}}(\mathbf{x}).
    \label{eq:matrix_equation}
\end{equation}
Here, we introduce the matrix operator $\hat{\mathbb{D}}(\mathbf{x})$ and the material parameters matrix $\hat{\mathbb{P}}_\omega(\mathbf{x})$ as
\begin{equation*}
    \hat{\mathbb{D}}(\mathbf{x}) =
    \begin{bmatrix}
        0                   & i\nabla_\mathbf{x} \cdot \\
        -i\nabla_\mathbf{x} & 0
    \end{bmatrix},
    \quad
    \hat{\mathbb{P}}_\omega(\mathbf{x}) =
    \begin{bmatrix}
        -\beta_\omega(\mathbf{x})   & 0 \\
        0                           & \rho_\omega(\mathbf{x})
    \end{bmatrix},
\end{equation*}
and acoustic field vector $\vec{\mathbb{F}}(\mathbf{x})$ and force source vector $\vec{\mathbb{J}}(\mathbf{x})$ as
\begin{equation*}
    \vec{\mathbb{F}}(\mathbf{x}) =
    \begin{bmatrix}
        p(\mathbf{x}) \\
        \mathbf{v}(\mathbf{x})
    \end{bmatrix},
    \quad 
    \vec{\mathbb{J}}(\mathbf{x}) =
    \begin{bmatrix}
        0 \\
        - i\mathbf{f}(\mathbf{x})
    \end{bmatrix}.
\end{equation*}

The solution of Eq.~\eqref{eq:matrix_equation} is given by the dyadic GF as follows:
\begin{equation}
    \vec{\mathbb{F}}(\mathbf{x}) = \int_V d\mathbf{x}^\prime \hat{\mathbb{G}}_\omega(\mathbf{x}, \mathbf{x}') \vec{\mathbb{J}}(\mathbf{x}^\prime).
    \label{eq:sol_with_gf}
\end{equation}
Here, the integral is taken over the whole system volume $V$ with all inhomogeneities and singularities of the material parameters matrix $\hat{\mathbb{P}}_\omega(\mathbf{x})$. Additionally, the GF satisfies the equation
\begin{equation}
    \left[ \omega \hat{\mathbb{P}}_\omega(\mathbf{x}) - \hat{\mathbb{D}}(\mathbf{x}) \right] \hat{\mathbb{G}}_\omega(\mathbf{x}, \mathbf{x}') = \hat{\mathbb{I}} \delta(\mathbf{x} - \mathbf{x}').
    \label{eq:green_function_dyadic_equation}
\end{equation}
If the GF is meromorphic, then it can be expanded in terms of its resonant poles as
\begin{equation}
    \hat{\mathbb{G}}_\omega(\mathbf{x}, \mathbf{x}^\prime) = \sum_n \frac{\vec{\mathbb{F}}_n(\mathbf{x}) \vec{\mathbb{F}}_n^\mathsf{T}(\mathbf{x}^\prime)}{\omega - \omega_n},
    \label{eq:green_func}
\end{equation}
where $\vec{\mathbb{F}}_n(\mathbf{x})$ and $\omega_n$ are the RSs and resonant frequencies satisfying Eq.~\eqref{eq:matrix_equation} without external sources:
\begin{equation}
    \hat{\mathbb{D}}(\mathbf{x}) \vec{\mathbb{F}}_n (\mathbf{x}) = \omega_n \hat{\mathbb{P}}_{\omega_n}(\mathbf{x}) \vec{\mathbb{F}}_n(\mathbf{x}).
    \label{eq:matrix_equation_hom}
\end{equation}
Obviously, the GF~\eqref{eq:green_func} has poles when the perturbation frequency matches the acoustic resonator eigenfrequencies $\omega = \omega_n$ in the complex $\omega$-plane. In general, however, the analytic structure of the Green's function may contain additional non-pole contributions. In particular, for the two-dimensional open acoustic system considered below, the Green's function possesses a branch cut in the complex-frequency plane. Therefore, the discrete set of RS poles alone does not constitute a complete spectral representation, and the branch-cut contribution must be included explicitly. The corresponding Green-function expansion, containing both the discrete resonant poles and the branch-cut contribution, is introduced in Eq.~\eqref{eq:green_function_2D}.

It is important to emphasize that RSs $\vec{\mathbb{F}}_n(\mathbf{x})$ of an open acoustic resonator are not normal modes of the Hermitian systems. They satisfy purely outgoing-wave boundary conditions and have complex eigenfrequencies $\omega_n$ with $\mathrm{Im}(\omega_n) \leq 0$ for the time dependence chosen as $\exp(-i\omega t )$. Since eigenfrequencies imaginary part $\mathrm{Im}(\omega_n) < 0$, this outgoing solution grows as $\exp(-\mathrm{Im}(\omega_n) r / c)$ at large distances when $r \rightarrow +\infty$. In fact, this exponential divergence is not a physical growth of the acoustic pressure $p$, but rather a consequence of the analytic continuation of the scattering problem to the complex-frequency plane. Physically, an RS describes the field remaining in the resonator in the absence of an external excitation. The temporal dependence of the RSs decays as $\exp\left[ \mathrm{Im}(\omega_n) t \right]$, while the spatially increasing outgoing tail at a fixed time can be interpreted in terms of radiation emitted at earlier retarded times $t-r/c$, when the amplitude of the decaying resonance was larger. Thus, an individual RS is a generalized eigenmode rather than a square-integrable physical field. A measurable acoustic field excited at a real frequency is obtained from a causal superposition of RSs and non-resonant background contributions, for which such divergences cancel. This is why the normalization of RSs requires a generalized bilinear form with surface terms, rather than the usual energy integral over the whole space~\cite{lalanne2018light, bochkarev2026electromagnetic}.

The RSs of Eq.~\eqref{eq:matrix_equation_hom} form an orthonormal and complete set of functions. We consider the normalization relation for the RSs is given as follows (see Sec.~I of the Supplemental material~\cite{Supplement_I} for details):
\begin{equation}
    \begin{aligned}
        \int_V d\mathbf{x} \vec{\mathbb{F}}_n^\mathsf{T} \frac{\partial}{\partial \omega} \left[ \omega \hat{\mathbb{P}}_\omega \right] \vec{\mathbb{F}}_n + \\
        \frac{1}{\omega_n} \int_S d\mathbf{S} \left[ \left(\mathbf{x} \cdot \nabla \vec{\mathbb{F}}_n\right)^\mathsf{T} \hat{\mathbb{D}} \vec{\mathbb{F}}_n - \vec{\mathbb{F}}_n^\mathsf{T} \hat{\mathbb{D}} \left(\mathbf{x} \cdot \nabla \vec{\mathbb{F}}_n\right) \right] = 1.
    \end{aligned}
    \label{eq:orthonormality}
\end{equation}
The first integral in Eq.~\eqref{eq:orthonormality} is taken over the entire system volume $V$ with all inhomogeneities, and the second one is taken over the closed surface $S$ on the boundaries of volume $V$. The surface contribution in Eq.~\eqref{eq:orthonormality} is essential for open systems. Since acoustic RSs obey outgoing boundary conditions at complex eigenfrequencies $\omega_n$, they generally diverge exponentially outside the resonator and cannot be normalized by the standard energy integral over all space, as noted above.

Although the acoustic equations and the normalization relation above are written in a general form that allows for frequency-dependent material parameters, in the following we restrict the RSE to nondispersive media and frequency-independent perturbations. Accordingly, we set $\rho_{\omega}(\mathbf{x}) \equiv \rho(\mathbf{x})$, $\beta_{\omega}(\mathbf{x}) \equiv \beta(\mathbf{x})$, and $\hat{\mathbb{P}}_{\omega}(\mathbf{x}) \equiv \hat{\mathbb{P}}(\mathbf{x})$, with $\Delta\hat{\mathbb{P}}_{\omega}(\mathbf{x}) \equiv \Delta\hat{\mathbb{P}}(\mathbf{x})$. Under this assumption, the perturbed eigenvalue problem remains linear in the complex eigenfrequency $\Omega$, and the perturbation matrix elements are independent of $\Omega$. The extension of the present formalism to dispersive acoustic media, for which the material operators must be evaluated at the perturbed eigenfrequency and the resulting eigenvalue problem becomes nonlinear, is beyond the scope of this work.

\subsection{Perturbative eigenvalue problem}
We now formulate how a perturbation of the material parameters modifies the RS eigenvalue problem. When the acoustic system is perturbed, i.e., 
$\hat{\mathbb{P}}'=\hat{\mathbb{P}}+\Delta\hat{\mathbb{P}}$, 
as shown in Fig.~\ref{Draft_Figure 1}, the eigenvalue problem~\eqref{eq:matrix_equation_hom} transforms into
\begin{equation}
    \hat{\mathbb{D}}(\mathbf{x})\vec{\mathbb{U}}(\mathbf{x})
    =
    \Omega
    \left[
        \hat{\mathbb{P}}
        +
        \Delta\hat{\mathbb{P}}
    \right]
    \vec{\mathbb{U}}(\mathbf{x}),
    \label{eq:matrix_equation_per}
\end{equation}
where $\vec{\mathbb{U}}(\mathbf{x})$ and $\Omega$ are the eigenvectors and complex eigenvalues of the perturbed system, respectively. In the most general linear constitutive description, the perturbation operator $\Delta\hat{\mathbb{P}}$ need not be block diagonal. Its off-diagonal components couple the pressure and particle-velocity fields and describe acoustic bianisotropy, commonly referred to as Willis coupling~\cite{muhlestein2017experimental, sieck2017origins, toftul2021directional}. In the following, however, we restrict our analysis to block-diagonal perturbations associated with spatial variations of the mass density and compressibility.

We obtain the solution of Eq.~\eqref{eq:matrix_equation_per} through the expansion of the perturbed system RSs $\vec{\mathbb{U}}(\mathbf{x})$ into the unperturbed system RS $\vec{\mathbb{F}}(\mathbf{x})$, using Eq.~\eqref{eq:green_func} in Eq.~\eqref{eq:sol_with_gf} and considering $\vec{\mathbb{J}}(\mathbf{x}) = - \Omega \Delta \hat{\mathbb{P}}(\mathbf{x}) \vec{\mathbb{U}}(\mathbf{x})$:
\begin{equation}
    \vec{\mathbb{U}}(\mathbf{x}) =  - \Omega \sum_n \frac{\vec{\mathbb{F}}_n(\mathbf{x}) }{\Omega - \omega_n} \int_V d\mathbf{x}' \vec{\mathbb{F}}_n^\mathsf{T}(\mathbf{x}') \Delta \hat{\mathbb{P}}(\mathbf{x}') \vec{\mathbb{U}}(\mathbf{x}').
    \label{eq:expansion}
\end{equation}

Let us expand the perturbed field $\vec{\mathbb{U}}(\mathbf{x})$ inside the system into the unperturbed RS $\vec{\mathbb{F}}_n(\mathbf{x})$ as
\begin{equation}
    \vec{\mathbb{U}}(\mathbf{x}) = \sum_n c_n \vec{\mathbb{F}}_n(\mathbf{x}),
    \label{eq:rse_expansion}
\end{equation}
With the previous expansion, Eq.~\eqref{eq:expansion} transforms to the following linear matrix form eigenvalue problem
\begin{equation}
    \sum_{n'} (\delta_{n n^\prime} + V_{n n^\prime}) c_{n^\prime} = \frac{\omega_n}{\Omega} c_n,
    \label{eq:rse_equation}
\end{equation}
where the matrix perturbation element $V_{nn'}$ is

\begin{equation}
    V_{n n^\prime} = \displaystyle \int_V d\mathbf{x}^\prime \vec{\mathbb{F}}_n^\mathsf{T}(\mathbf{x}^\prime) \Delta \hat{\mathbb{P}}(\mathbf{x}^\prime) \vec{\mathbb{F}}_{n^\prime}(\mathbf{x}^\prime).
    \label{eq:matrix_general}
\end{equation}
The structure of Eq.~\eqref{eq:rse_equation} corresponds to the Brillouin-Wigner form of perturbation theory in both quantum mechanics~\cite{bang1978expansion,lind1993completeness} and electrodynamics~\cite{muljarov2011brillouin}, rather than to the conventional Rayleigh-Schrödinger perturbation expansion~\cite{landau1991quantum}. It has a direct physical meaning, because it measures the overlap between the spatial profile of the perturbation $\Delta \hat{\mathbb{P}}(\mathbf{x}^\prime)$ and two RSs $\vec{\mathbb{F}}_n(\mathbf{x})$ of the unperturbed system. In brief, $V_{nn'}$ plays the same role as a perturbation matrix element in quantum mechanics. Diagonal elements of Eq.~\eqref{eq:matrix_general} with $n = n'$ describe the self-action of a mode and determine, to first order, the shift of its complex eigenfrequency $\omega_n$, whereas off-diagonal elements of Eq.~\eqref{eq:matrix_general} with $n \neq n'$ describe the perturbation-induced coupling between different RSs. In a Hermitian quantum-mechanical problem, such off-diagonal matrix elements are directly related to transition amplitudes, and transition probabilities are proportional to $|V_{nn'}|^2$. Instead, in a non-Hermitian acoustic problem, the matrix element $V_{nn'}$ is a complex coupling amplitude between leaky RSs, responsible for modal hybridization, avoided crossings, resonance splitting, and changes of the radiative linewidth. In the RSE framework, the RS $\vec{\mathbb{U}}_n(\mathbf{x})$ of the perturbed system is instead expanded over the whole basis of unperturbed RS $\vec{\mathbb{F}}_n(\mathbf{x})$ with Eq.~\eqref{eq:rse_expansion}, and the resulting matrix problem given by Eq.~\eqref{eq:rse_equation} is solved directly.

One can reformulate Eq.~\eqref{eq:rse_equation} as a matrix eigenvalue problem of the form $\hat W \mathbf{c} = \Omega \mathbf{c}$, where the matrix elements are given by $W_{nn'} = \omega_n (\delta_{nn'} + V_{nn'})^{-1}$, and the RSE coefficients $c_n$ form its eigenvector $\mathbf{c}$. Diagonalization of the matrix $\hat W$ yields the eigenfrequencies $\Omega$ as well as the corresponding expansion coefficients $c_n$ for the perturbed RS $\vec{\mathbb{U}}_n(\mathbf{x})$.

\subsection{Normalization of perturbed resonant states}
Since the RSE framework allows us to expand the perturbed field into an unperturbed series, each pressure field RS will contribute to the perturbed state. Consequently, the perturbed RS field needs to be renormalized numerically with Eq.~\eqref{eq:orthonormality}. Let us introduce functional $N[\vec{\mathbb{F}}_n] \sim \vec{\mathbb{F}}_n^2$ formed by the both integrals from Eq.~\eqref{eq:orthonormality}. Substituting Eq.~\eqref{eq:rse_expansion}, one can obtain the following form of the functional $N[\vec{\mathbb{F}}_n]$
\begin{equation}
    \mathcal{N}[\vec{\mathbb{F}}_n] = \sum_{n,n'} c_n c_{n'} \hat{\mathbb{M}}_{nn'}(\Omega).
    \label{eq:normalization_functional}
\end{equation}
Here, $\hat{\mathbb{M}}_{nn'}(\Omega)$ is the normalization matrix element, defined by both volume and surface contributions from Eq.~\eqref{eq:orthonormality}. If the eigenvector $\mathbf{c}$ of $\hat W \mathbf{c} = \Omega \mathbf{c}$ is found, we can define the value of the normalization functional Eq.~\eqref{eq:normalization_functional} of the RS, written in the basis of unperturbed eigenstates as $\mathcal{N} = \mathbf{c}^\mathsf{T} \hat{M} \mathbf{c}$. Perturbed system acoustic field vector is defined with the normalization as
\begin{equation*}
    \vec{\mathbb{U}}(\mathbf{x}) = \frac{1}{\sqrt{\mathbf{c}^\mathsf{T} \hat{\mathbb{M}} \mathbf{c}}} \sum_n c_n \vec{\mathbb{F}}_n(\mathbf{x}).
\end{equation*}

\section{Resonant-State Expansion for a 2D Cylindrical Resonator}
\label{sec:application}
\subsection{Reference cylindrical resonator}
We first consider the source-free acoustic problem, $\mathbf{f}(\mathbf{x}) = 0$, for which the RS are defined as the eigenmodes of Eqs.~\eqref{eq:eq_system_complex}. Eliminating the velocity field from these equations by substituting the second equation into the first one, one obtains the Helmholtz equation $\nabla \cdot [\rho^{-1}(\mathbf{x}) \nabla p(\mathbf{x})] + \omega^2 \beta(\mathbf{x}) p(\mathbf{x}) = 0$ for the pressure field $p(\mathbf{x})$, with sound speed $c(\mathbf{x}) = [\beta(\mathbf{x}) \rho(\mathbf{x})]^{-1/2}$, which in general depends on coordinate $\mathbf{x}$. Once the pressure eigenmodes are found, the corresponding velocity field $\mathbf{v}(\mathbf{x})$ is obtained from the second of Eqs.~\eqref{eq:eq_system_complex}. These pressure and velocity fields together form the RSs of the acoustic system.

\begin{figure}[t]
    \centering
    \includegraphics[width = \linewidth]{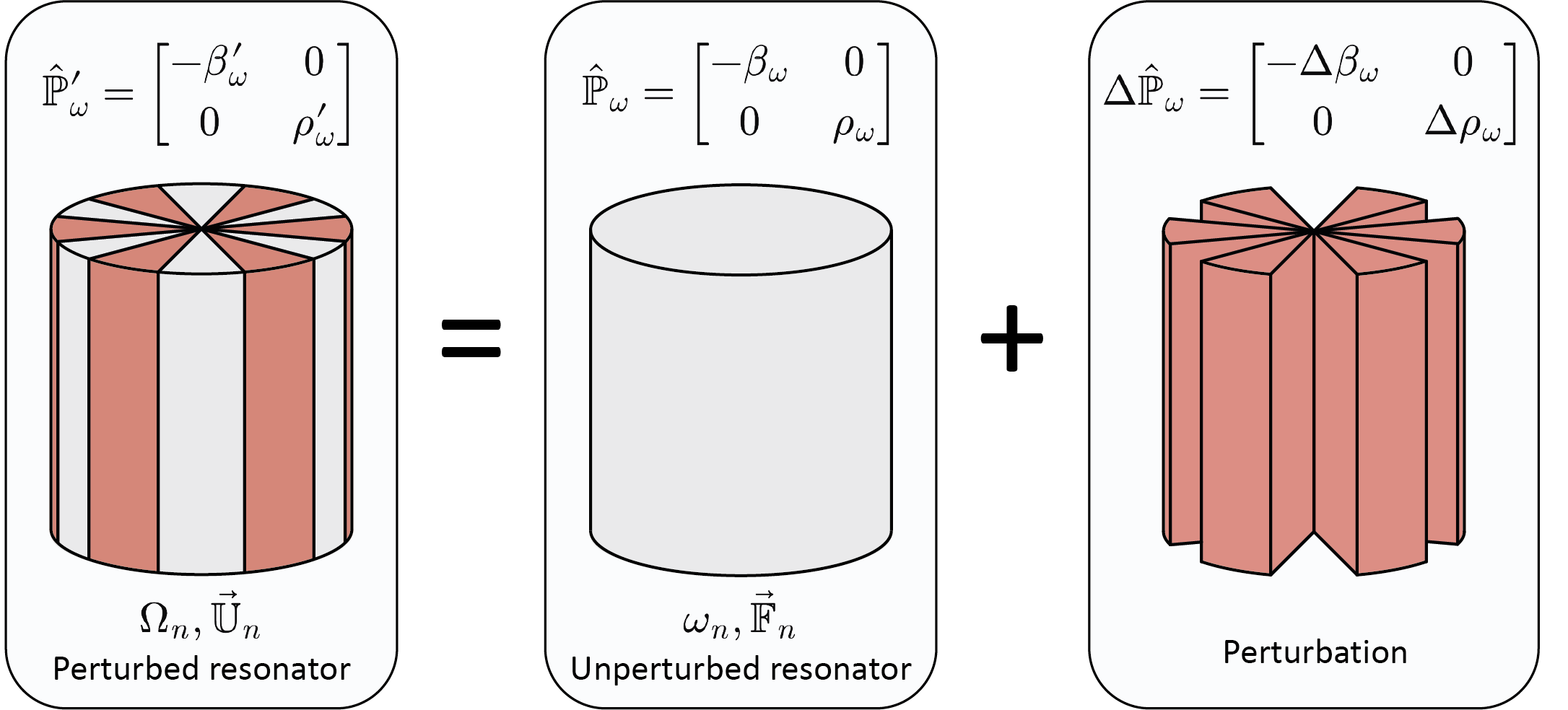}
    \caption{Schematic representation of the RSE framework. The perturbed resonator, characterized by the material-parameter operator $\hat{\mathbb{P}}'_\omega$, eigenfrequencies $\Omega_n$, and eigenvectors $\vec{\mathbb{U}}_n$, is represented as the combination of an analytically solvable unperturbed resonator, described by $\hat{\mathbb{P}}_\omega$, $\omega_n$, and $\vec{\mathbb{F}}_n$, and a spatial perturbation of the material parameters $\Delta\hat{\mathbb{P}}_\omega$.}
    \label{Draft_Figure 1}
\end{figure}

Figure~\ref{Draft_Figure 1} illustrates the main concept of the RSE framework. Various inclusions, such as porous materials or metamaterial absorbers, may be present within the system and influence the eigenfrequencies and eigenfunctions. This significantly increases the complexity of the problem and often renders it analytically intractable. The solution to this challenge is provided by the RSE, which allows one to express the solution of a complex system with $\{ \Omega_n, \vec{\mathbb{U}}_n(\mathbf{x}) \}$ in terms of an analytically solvable reference system, characterized by $\{ \omega_n, \vec{\mathbb{F}}_n(\mathbf{x}) \}$, with an added perturbation $\Delta \hat{\mathbb{P}}$. In this work, three types of perturbations are considered for both analytical and numerical investigation in order to develop an acoustic RSE framework and to validate its accuracy against other simulation methods. Firstly, in the case of homogeneous perturbation, the initial unperturbed resonator is replaced with another material, whose parameters are constant throughout the entire internal space of the resonator. This represents the simplest type of perturbation and is used to demonstrate the accuracy of the RSE method. Next, the radial perturbation of the material parameters is considered. In this case, the perturbation is angularly independent and changes with respect to radius. That perturbation modifies the radial distribution of the resonator eigenmodes, whereas axial symmetry remains unchanged. Finally, we consider sectoral perturbation, which breaks the axial symmetry $C_\infty$ of the initial cylindrical resonator. In this case, the resonator is separated into different sectors with their own material parameters. The analysis of these types of perturbations within the RSE framework enables the investigation of wave scattering by complex structures with arbitrary spatial variations of material parameters.

As a starting point, we choose a two-dimensional analytically solvable reference system. Specifically, a homogeneous circular cylinder of radius $R = 10~\mathrm{cm}$, embedded in a homogeneous background medium (air with sound velocity $c_b = 343~\mathrm{m/s}$ and density $\rho_b = 1.2~\mathrm{kg/m}^3$). The cylinder material parameters are normalized relative to the background air parameters, $\rho_0 = 10\rho_b$ and $c_0 = c_b/2$. Here, we neglect the frequency dependence of material parameters and choose these specific values, which are corroborated by numerical analyses of experimentally acquired data for analogous resonant systems, specifically, labyrinth-like and coil structures, operating at comparable frequencies~\cite{liang2013_space-coiling,maurya2016double,yin2026design}. Hence, compressibility for both materials is $\beta = 1/\rho c^2$.

This system is used as the unperturbed basis of the RSE. More general two-dimensional acoustic systems, such as cylinders with radial or azimuthal inhomogeneity, are then treated as perturbations of this reference configuration. Since such inhomogeneous systems generally do not admit closed-form analytical solutions, the RSE provides an efficient and systematic way to determine their resonances.

\begin{figure}[t]
    \centering
    \includegraphics[width = \linewidth]{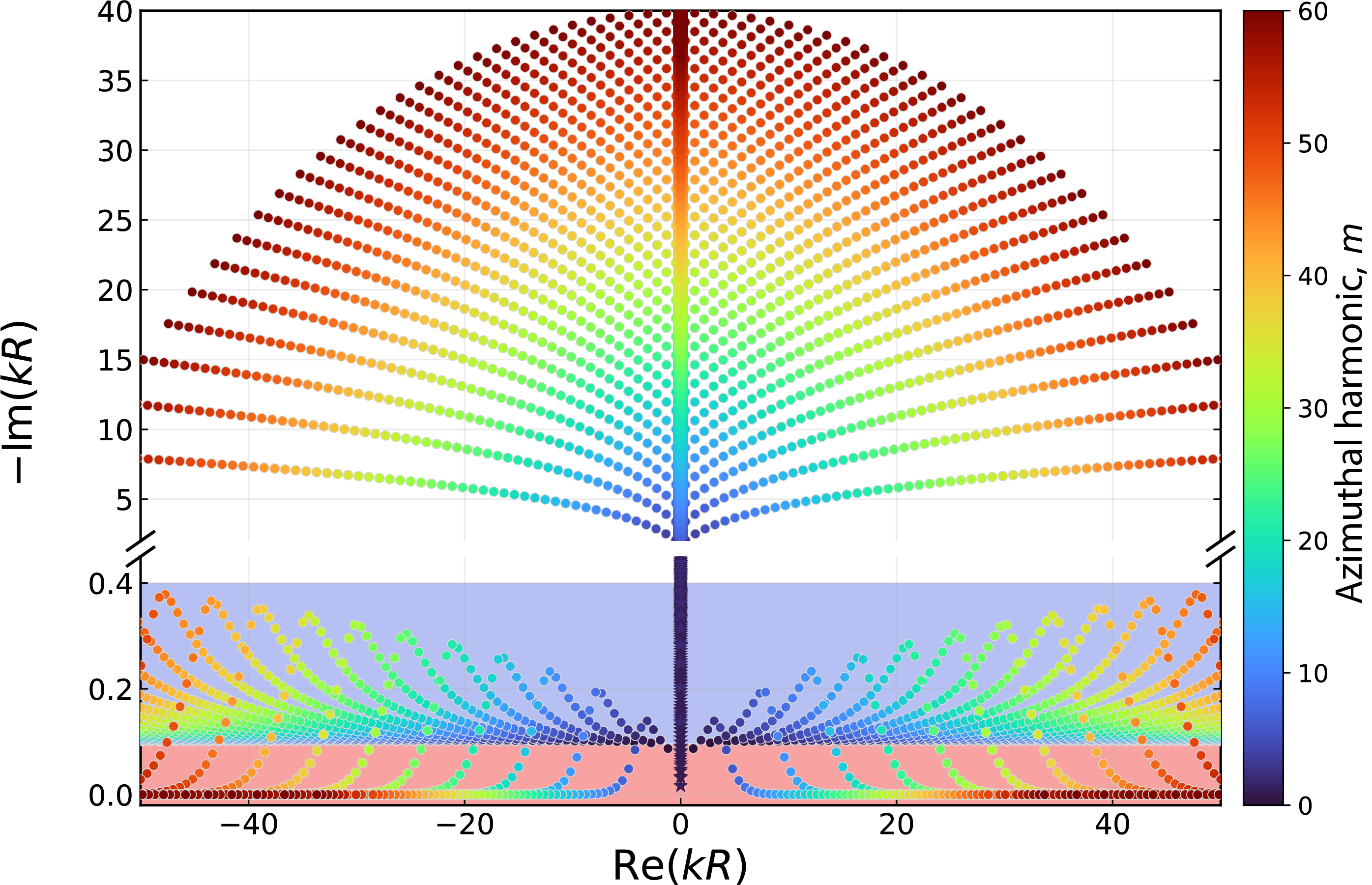}
    \caption{Eigenfrequency spectra of the unperturbed system are shown for azimuthal numbers $m = 0~\mathrm{to}~60$ consisting of regular modes (coloured circles), and branch cut modes with $\mathrm{Re}(kR) = 0$ (coloured stars). The considered cylindrical structure consists of the open resonator with radius $R$ and material parameters $\rho_0 = 10 \rho_b, c_0 = c_b/2$ normalized to the background air material parameters $\rho = 1.2~\mathrm{kg/m}^3, c = 343~\mathrm{m/s}$. Compressibility for both materials is defined by $\beta = 1/\rho c^2$.}
    \label{Draft_Figure 2}
\end{figure}

For this cylindrically symmetric system, the Helmholtz equation takes the form
\begin{equation}
    \left[ \frac{1}{r} \frac{\partial}{\partial r}\left( r \frac{\partial}{\partial r} \right) + \frac{1}{r^2} \frac{\partial^2}{\partial \varphi^2} + \frac{\omega_n^2}{c^2} \right] p_n(r, \varphi) = 0,
    \label{eq:cylindrical_eq}
\end{equation}
where all material parameters including mass density $\rho$, compressibility $\beta$ and sound velocity $c$ are azimuthally homogeneous and depend only on the radial coordinate:
\begin{equation}
    \{\beta, \rho, c\} =
    \begin{cases}
        \beta_0, \rho_0, c_0, \quad r \leq R, \\
        \beta_b, \rho_b, c_b, \quad r > R.
    \end{cases}
    \label{eq:parameter}
\end{equation}

Due to the cylindrical symmetry, the azimuthal index $m$ is a good quantum number that takes integer values, giving the number of field oscillations around the cylinder. The RSs of the unperturbed system can therefore be written as:
\begin{equation}
    p_{nm}(r, \varphi) = R_m(r, \omega_n) \chi_m(\varphi).
    \label{eq:res_states}
\end{equation}
Here $\chi_m(\varphi)$ representing angular part of RS~\eqref{eq:res_states} is defined by
\begin{equation}
    \chi_m(\varphi) =
    \begin{cases}
        \displaystyle \frac{1}{\sqrt{\pi}} \sin(m\varphi), m < 0, \\
        \displaystyle \frac{1}{\sqrt{2\pi}}, m = 0, \\
        \displaystyle \frac{1}{\sqrt{\pi}} \cos(m\varphi), m > 0,
    \end{cases}
    \label{eq:Angular_part}
\end{equation}
and satisfies the orthonormality relation
\begin{equation}
    \int_0^{2\pi} d\varphi \chi_m(\varphi) \chi_{m'}(\varphi) = \delta_{mm'}.
    \label{eq:angular_part_orthonormality}
\end{equation}
The radial part of Eq.~\eqref{eq:res_states} has the form
\begin{equation}
    R_m(r, \omega_n) = A_n^m
    \begin{cases}
        J_m(k_n r) / J_m(k_n R), r \leq R \\
        H_m(k_n^{ex} r) / H_m(k_n^{ex} R), r > R
    \end{cases},
    \label{eq:Radial_part}
\end{equation}
where $J_m(k_nr)$ and $H_m(k_n^{ex} r)$ are cylindrical $m$-order Bessel functions and Hankel functions of the first kind, respectively, $k_n = \omega_n/c_0$ and $k_n^{ex} = \omega_n/c_b$. It follows from Eqs.~\eqref{eq:eq_system_complex} and \eqref{eq:res_states} that the velocity components corresponding to the pressure eigenmode can be written in cylindrical coordinates as
\begin{equation*}
    \begin{aligned}
        v_r(r,\varphi)
        &= -\frac{i}{\omega_n \rho(r,\varphi)}
        R_m'(r,\omega_n)\chi_m(\varphi), \\
        v_\varphi(r,\varphi)
        &= -\frac{i}{\omega_n \rho(r,\varphi)}
        \frac{1}{r}R_m(r,\omega_n)\chi_m'(\varphi).
    \end{aligned}
\end{equation*}
The normalization constant for RSs in cylindrical coordinates for the non-dispersive materials, for which $\partial \hat{\mathbb{P}} / \partial \omega  = 0$, in accordance with the orthonormality relation~\eqref{eq:orthonormality}, takes the form:


\begin{equation}
    \begin{aligned}
        1 ={}& - 2 \int_0^{2\pi} d\varphi \int_0^R dr\, r\, \beta(r,\varphi) p_n^2(r,\varphi) \\
        &+ \frac{R}{k_n^2} \int_0^{2\pi} d\varphi\, \beta(R^+,\varphi)
        \left[
            r \left(\frac{\partial p_n}{\partial r}\right)^2
            - r p_n \frac{\partial^2 p_n}{\partial r^2}
        \right]_{r=R^+} \\
        &- \frac{R}{k_n^2} \int_0^{2\pi} d\varphi\, \beta(R^+,\varphi)
        \left[
            p_n \frac{\partial p_n}{\partial r}
        \right]_{r=R^+}.
    \end{aligned}
    \label{eq:orthonormality_cyl}
\end{equation}

We find that the eigenfunctions are normalized according to~\eqref{eq:orthonormality_cyl} with the normalization constant
\begin{equation}
    \begin{aligned}
        A_n^m = \frac{1}{R} \Bigg\{ \beta_b - \beta_0
        + \left( \frac{\rho_b}{\rho_0} - 1 \right)
        \Bigg[ &\beta_0 \left(\frac{J'_m(k_nR)}{J_m(k_nR)}\right)^2 \\
        &+ \frac{m^2}{\rho_b \omega_n^2 R^2} \Bigg]
        \Bigg\}^{-1/2}.
    \end{aligned}
    \label{eq:norm_const}
\end{equation}
The general form of the constant normalization for a homogeneous cylinder in a medium is simplified in two specific cases. First, for azimuthally homogeneous modes $m = 0$, the last term is eliminated. Second, if the resonator density $\rho_0$ and the background medium density $\rho_b$ are equal, regardless of the azimuthal moment, the normalization constant is simplified to $A_n^m = \left( R \sqrt{\beta_b - \beta_0} \right) ^{-1}$. Once $\beta_b = \beta_0$, physically, a resonator does not exist, as well as a complete and orthogonal set of RSs; meanwhile, the pressure field can not be normalized in open space.

The boundary condition at the surface of the cylindrical resonator, specifically the continuity of the pressure derivative normalized to the corresponding density, yields the following secular equation for the RS frequencies
\begin{equation}
    \begin{aligned}
        D_m(\omega_n) ={}& \gamma J'_m(k_nR) H_m(k_n^{ex}R) \\
        &- H'_m(k_n^{ex}R) J_m(k_nR) = 0 .
    \end{aligned}
    \label{eq:boundary_cond}
\end{equation}
where $\gamma = \rho_b c_b / \rho_0 c_0$. We find the solution of Eq.~\eqref{eq:boundary_cond} numerically for the eigenmodes with azimuthal number $m = 0~\mathrm{to}~60$. Eigenstates characterized by a real part approaching zero and a large imaginary component correspond to evanescent waves outgoing from the resonator area (shown in Fig.~\ref{Draft_Figure 2} by shading with white colour). In contrast, eigenfrequencies with a small imaginary part are associated with high-order RSs, ''whispering gallery''-like modes, and generally localized inside the considered resonator. These modes exhibit extremely high quality factors $Q$ (shaded in red colour in Fig.~\ref{Draft_Figure 2}). Leaky modes of the resonator, characterized by a pressure field distributed throughout the entire considered area, exhibit significantly higher energy losses to the surrounding environment when compared to the "whispering gallery"-like modes, which are illustrated in Figure~\ref{Draft_Figure 2} and shaded in blue. To accelerate the simulation and improve the accuracy of eigenvalue calculations, we employ Cauchy's argument principle~\cite{xu2023argument, nagarsheth2021somenew} paired with the Newton method to solve the transcendental equation~\eqref{eq:boundary_cond}. 

\subsection{Branch-cut contribution}

In addition to the discrete resonant poles, the Green's function may contain non-pole singularities that must be included in its spectral representation. In the two-dimensional acoustic problem considered here, the Green's function possesses a branch cut along the imaginary-frequency axis. Its spectral representation therefore consists of a sum over the discrete RS poles supplemented by an integral along the branch cut~\cite{doost2013resonant},
\begin{equation}
    \begin{aligned}
        G_\omega(r,r') ={}& \sum_n
        \frac{R_m(r,\omega_n)R_m(r',\omega_n)}
        {2\omega(\omega-\omega_n)}
        \\
        &+ \int_{-i\infty}^{0} d\omega'\,
        \frac{R_m(r,\omega')R_m(r',\omega')}
        {2\omega(\omega-\omega')}
        \sigma_m(\omega'),
    \end{aligned}
    \label{eq:green_function_2D}
\end{equation}
where $\sigma_m(\omega)$ is the branch-cut weight function, which determines the contribution of the continuum along the cut and is given by
\begin{equation}
    \sigma_m(\omega) =
    \frac{J_m^2(kR)}
    {\pi^2 k A_m^2(\omega) D_m^+(\omega) D_m^-(\omega)}.
    \label{eq:sigma_function}
\end{equation}
Considered two-dimensional GF~\eqref{eq:green_function_2D} corresponds to the dyadic GF~\eqref{eq:green_func}, satisfying the Eq.~\eqref{eq:green_function_dyadic_equation} with the factor $2 \omega$. We discuss the details of the two-dimensional acoustic problem GF in Sec.~II and~III of the Supplemental material~\cite{Supplement_I}. The above-noted weight function $\sigma_m(\omega)$ introduces cut-pole contributions to the eigenfrequencies' positions in the perturbed system spectra. For the low azimuthal numbers, its contribution is insubstantial, as shown in~\cite{Supplement_I}, unlike the high azimuthal numbers, where one needs to take into account a large number of cut poles, leading to long-time simulations.

Before proceeding to particular perturbation profiles, we introduce the perturbation matrix element in its most general pressure-only form. Using Eq.~\eqref{eq:matrix_general} and eliminating the velocity field via Eq.~\eqref{eq:eq_system_complex}, the perturbation matrix elements can be expressed solely in terms of the pressure eigenfunctions of the unperturbed system as follows:
\begin{equation}
    \begin{aligned}
        V_{nn'} = - \int_V d\mathbf{x} \Delta \beta(\mathbf{x})\, p_n(\mathbf{x}) p_{n'}(\mathbf{x}) \\
        - \frac{1}{\omega_n \omega_{n'} \rho_0^2} \int_V d\mathbf{x} \Delta \rho(\mathbf{x})
    \nabla p_n(\mathbf{x}) \nabla p_{n'}(\mathbf{x})        
    \end{aligned}
    \label{eq:general_V_pressure}
\end{equation}
with the integration restricted to the region where the material parameters are perturbed. This form clearly separates the effects of compressibility and density variations and provides a unified starting point for treating homogeneous, radial, and sectoral perturbations considered below (see details in Sec.~IV of the Supplemental material~\cite{Supplement_I}).

\subsection{Homogeneous material perturbation}

\begin{figure}[t]
    \centering
    \includegraphics[width = \linewidth]{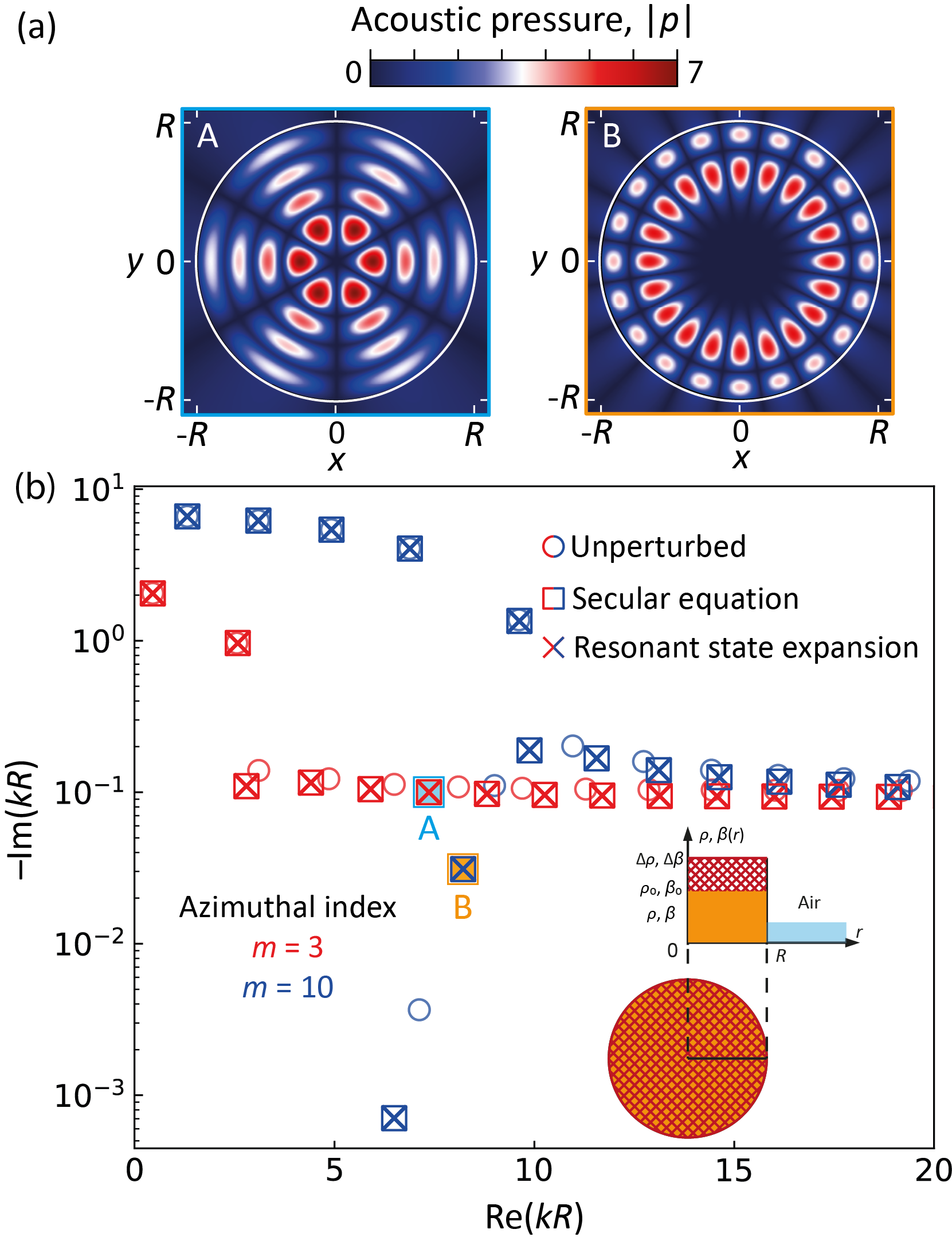}
    \caption{(a) Pressure field module distributions for two modes with different pairs of azimuthal and radial indices $(m, \ell)$: field A has $m = 3, \ell = 4$, and field B has $m = 10, \ell = 2$. The cylindrical resonator border is shown by the white circle in both pictures. (b) Eigenfrequency spectra of the unperturbed (open circles) and homogeneously perturbed system obtained by numerical simulation with Eq.~\eqref{eq:boundary_cond} (open squares) and RSE (crosses) are presented for azimuthal numbers $m = 3$ (red colour) and $m = 10$ (blue colour). The square contours show frequency positions in spectra for specific pressure field distributions. The inset figure displays the perturbed system and the material parameters constant profile for both systems.}
    \label{Draft_Figure 3}
\end{figure}

First and foremost, we begin with the simplest perturbation of the reference system: a homogeneous modification of the material parameters inside the cylinder (inset figure in panel (c) in Fig.~\ref{Draft_Figure 3}). Such a perturbation is independent of the polar angle and can be written as $\Delta \rho(r) = \Delta \rho \ \theta(R-r)$ and $\Delta\beta(r) = \Delta\beta \ \theta(R-r)$. Since this perturbation preserves the cylindrical symmetry of the unperturbed resonator, the azimuthal index $m$ does not couple. According to the orthonormality of the angular functions $\chi_m(\varphi)$, given by Eq.~\eqref{eq:angular_part_orthonormality}, the perturbation matrix is diagonal in the azimuthal basis, i.e. $V_{mm'}^{\ell\ell'} = V_m^{\ell\ell'}\delta_{mm'}$, and only RSs with the same azimuthal number $m$ are coupled. The perturbation matrix element~\eqref{eq:matrix_general} has a block-diagonal structure $V = \mathrm{diag} \left\{ V_{m = 0}, V_{m = 1}, \ldots, V_m \right\}$, where each matrix block $\hat{V}^{(m)}$ may still contain off-diagonal elements with respect to the radial indices $\ell$ and $\ell'$ define field distribution along the cylindrical resonator radius, but different azimuthal numbers $m$ and $m'$ do not couple.

We choose the perturbations in the material parameters to be related to those of the unperturbed resonator, with $\Delta \rho = 0.1 \rho_0$ and $\Delta \beta = 0.1 \beta_0$. Within the RSE framework, matrix perturbation elements are determined analytically for the non-diagonal case $n \neq n'$ and the diagonal case $n = n'$ (see Supplemental Material~\cite{Supplement_I}, Sec.~IV~A). Importantly, since the homogeneous perturbation does not change the circular symmetry of the unperturbed problem, the secular equation~\eqref{eq:boundary_cond} with parameter $\gamma$ changed, still gives us the RSs of the perturbed system. To illustrate the accuracy of the RSE, we compare numerically acquired results for the perturbed system, obtained via Eq.~\eqref{eq:boundary_cond}, with the RSE calculations discussed above, as shown in Fig.~\ref{Draft_Figure 3}. In addition, for reference, we also include the unperturbed spectra for specific azimuthal numbers $m = 3$ and $m = 10$.

Figure~\ref{Draft_Figure 3}(a) demonstrate the spatial distributions of the pressure field modulus for specific resonator eigenmodes with $m = 3$ and $m = 10$. In both cases, the field distribution structure exhibits symmetry with respect to a specific value of the azimuthal number $m$, and displays twice the number of local field maxima compared to the $m$ value. The degree of radial degeneration depends on the radial quantum number $\ell$, which determines the number of local maxima along the cylinder radius. For example, the field A with $m = 3$, $\ell$ equals 4 shown in Fig.~\ref{Draft_Figure 3}(a). In contrast, the field B in Fig.~\ref{Draft_Figure 3}(a) with $m = 10$ has the radial index $\ell$ equals 2.

The homogeneous perturbation is especially useful because it provides an analytically verifiable test case for the acoustic RSE. In this situation, the perturbed resonant frequencies can be determined independently from the exact secular equation of a homogeneous cylinder with modified material parameters. Therefore, the RSE results can be benchmarked directly against the exact solution, without the need for additional numerical reference calculations.

\subsection{Radially varying material perturbation}

\begin{figure}[t]
    \centering
    \includegraphics[width = \linewidth]{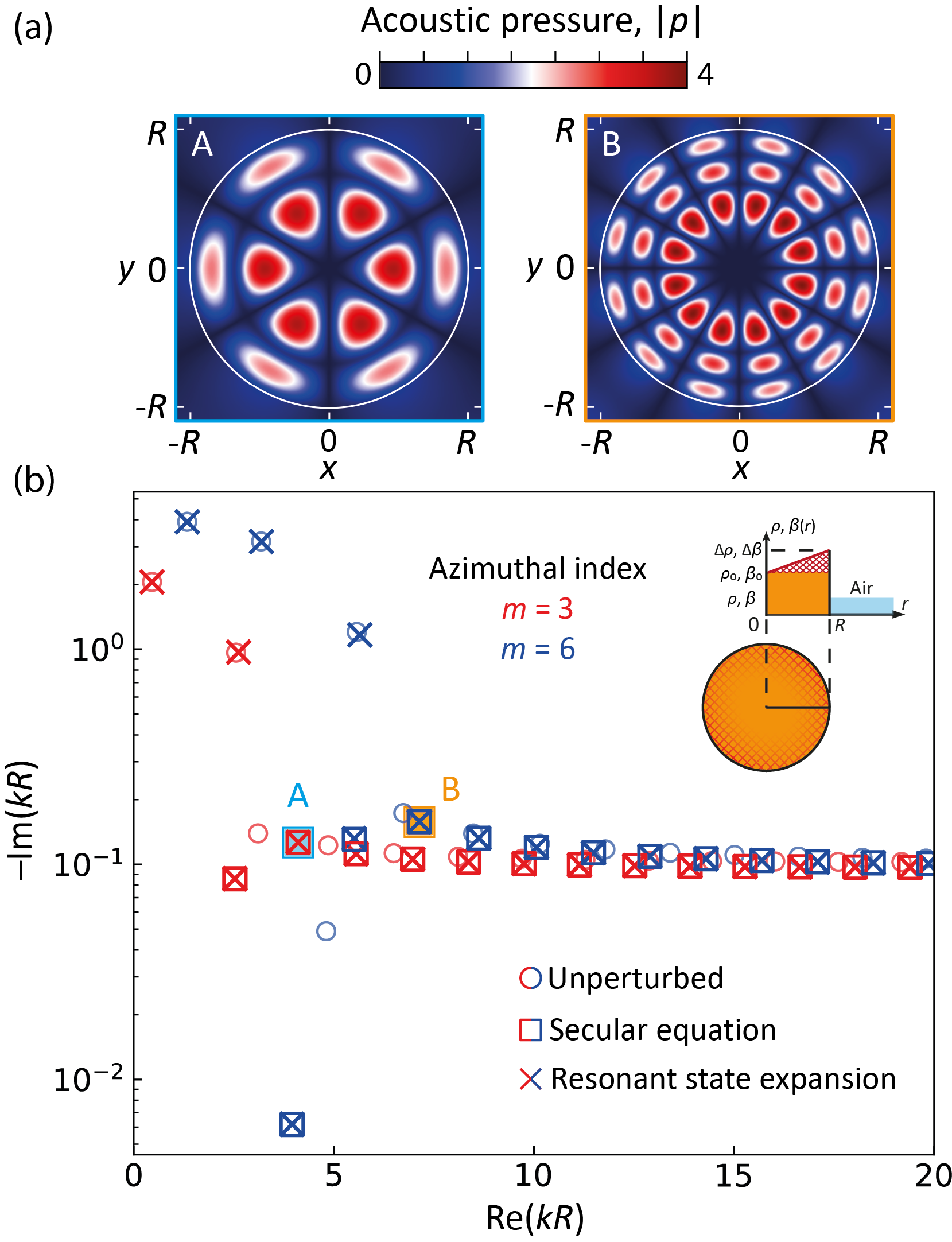}
    \caption{(a) Pressure field module distributions for azimuthal index $m = 3$ (A) and $m = 6$ (B). The cylindrical resonator border is shown by the white circle in both pictures. (b) Eigenfrequency spectra of the unperturbed (open circles) and radially perturbed system obtained by numerical simulation (open squares) and RSE (crosses) are presented for azimuthal numbers $m = 3$ (red colour) and $m = 6$ (blue colour). The square contours show frequency positions in spectra for specific pressure field distributions. The inset figure displays the perturbed system and the material parameters linear profile for both systems.}
    \label{Draft_Figure 4}
\end{figure}

Next, we consider a radial perturbation of the cylinder material parameters, preserving cylindrical symmetry. In this scenario, the perturbation depends solely on the radial coordinate, altering the material uniformly in all directions from the axis. This resembles an isotropic perturbation, as shown in Fig.~\ref{Draft_Figure 4}. The orthogonality of the angular functions $\chi_m(\varphi)$ ensures that each azimuthal index $m$ remains unaffected by the others and stays separated.

We introduce this angularly isotropic radial perturbation as $\Delta\rho(r) = \Delta\rho f_\rho(r) \theta(R-r)$ and $\Delta\beta(r) = \Delta\beta f_\beta(r) \theta(R-r)$. Here, $f_\rho(r)$ and $f_\beta(r)$ are dimensionless functions. They describe the dependence of the material parameters on the distance from the resonator center. We discuss the details of RSE for radial perturbation of various forms in Supplemental Material~\cite{Supplement_I}, Sec.~IV~B. In general, the integral~\eqref{eq:general_V_pressure} is taken for arbitrary dimensionless functions $f_\mathcal{\rho}(r)$ and $f_\mathcal{\beta}(r)$, dividing into two parts. However, in the absence of radial dependence, i.e., $f_\mathcal{\rho}(r) = f_\mathcal{\beta}(r) = \mathrm{const.}$, the integrals simplify to the form of the homogeneous perturbation. Finally, as noted, they are taken for diagonal $n = n'$ and non-diagonal $n \neq n'$ cases.

To be specific, we define radial characteristic functions as $f_\rho(r) = f_\beta(r) = r/R$ for numerical simulations. This leads to a smooth linear variation of material parameters along the cylinder radius, transitioning from $\{ \rho_0, \beta_0 \}$ in the center of the structure to $\{ \rho_0 + \Delta\rho, \beta_0 + \Delta\beta \}$ at the cylinder boundary. Likewise to the homogeneous perturbation, for the numerical analysis, we choose a perturbation of material parameters related to the material parameters of the initial resonator as $\Delta \rho = 0.2 \rho_0$ and $\Delta \beta = 0.4 \beta_0$.

In order to validate the accuracy of RSE for the considered perturbation form, we numerically simulate an open cylindrical resonator with radial dependence of the material parameters. Numerical computations were performed in COMSOL Multiphysics using the finite-element method with a user-controlled mesh and an extra-fine element size. Initially, to accelerate calculations, the one-dimensional axisymmetric model was used. The cylindrical wave radiation condition was used in order to simulate open boundary conditions. The Pressure acoustics module was used in eigenfrequency calculation for Fig.~\ref{Draft_Figure 4}.

\begin{figure*}[t]
    \centering
    \includegraphics[width = \linewidth]{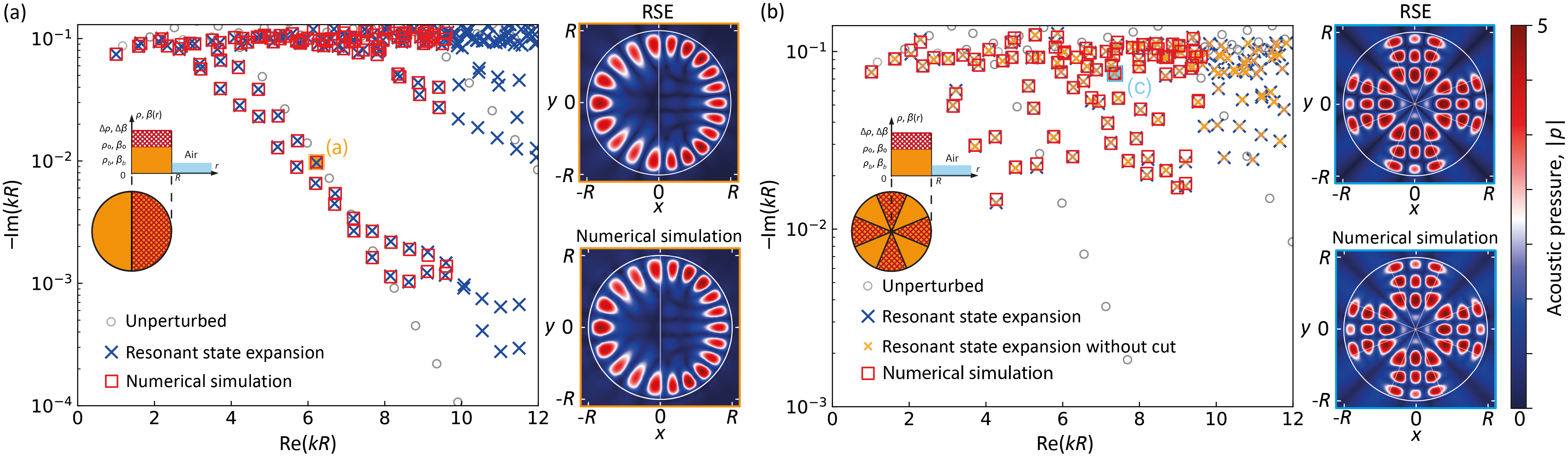}
    \caption{Comparison between pressure field module distributions, obtained by numerical simulations and RSE, is conducted for the selected mode (a) within the system with half-perturbation, highlighted in orange in (b), and for the selected mode (c) within the $C_4$-symmetric system, highlighted in blue in (d). The inset images illustrate the perturbed configurations alongside the sectoral profiles of the material parameters for both systems.}
    \label{Draft_Figure 5}
\end{figure*}

As shown in Fig.~\ref{Draft_Figure 4}, the RSE results are in close agreement with the numerical simulations, while requiring substantially less computational time by expanding the perturbed states in the basis of the unperturbed resonator. The contribution of cut poles to the eigenfrequency spectra remains significant due to the high-$Q$ factor modes analysis. Considered radial perturbation changes material parameter values close to the cylinder border and changes the radial modal profile while preserving its azimuthal order, as shown in panel (a) in Fig.~\ref{Draft_Figure 4}, in general.

\subsection{Sectoral symmetry-breaking perturbation}

Finally, we introduce a radially independent perturbation with a nontrivial angular dependence,
$\Delta \rho(r,\varphi) = \Delta\rho S_\rho(\varphi)\theta(R-r)$ and
$\Delta \beta(r,\varphi) = \Delta\beta S_\beta(\varphi)\theta(R-r)$.
Unlike the homogeneous and radial perturbations considered above, this perturbation explicitly reduces the rotational symmetry of the unperturbed cylindrical resonator from $C_\infty$ to a discrete symmetry $C_N$. Here, the dimensionless characteristic functions $S_\rho(\varphi)$ and $S_\beta(\varphi)$ describe the angular distribution of the material perturbation and are equal to 1 inside the corresponding sectoral regions (for details see Supplemental Material~\cite{Supplement_I}, Sec.~IV~C). In the following, we consider identical even profiles, $S_\rho(\varphi)=S_\rho(-\varphi)$ and $S_\beta(\varphi)=S_\beta(-\varphi)$, characterized by a rotational symmetry $C_N$, where $N$ is the order of the rotational axis and the corresponding rotation angle is $\varphi_N=2\pi/N$. Such discrete rotational symmetries are common in sectoral and space-coiling acoustic meta-atoms.

We consider two representative perturbation profiles corresponding to $N=1$ and $N=4$, i.e., perturbations with $C_1$ and $C_4$ symmetry, respectively, as shown in Fig.~\ref{Draft_Figure 5}. Analytical evaluation of the angular part of the perturbation matrix element~\eqref{eq:general_V_pressure} shows that the rotational symmetry imposes selection rules on the coupling between different azimuthal modes. As a result, the perturbation matrix acquires a block structure in which certain coupling terms vanish depending on the symmetry of the angular profiles $S_\rho(\varphi)$ and $S_\beta(\varphi)$. The corresponding derivation and explicit expressions for these symmetry-dependent coupling terms are given in Supplemental Material~\cite{Supplement_I}, Sec.~IV~C.

For numerical analysis, we set the material parameter perturbations as $\Delta \rho = 0.2 \rho_0$ and $\Delta \beta = 0.4 \beta_0$. The half-cylinder perturbation occupies an interval of $\varphi = 180^{\circ}$, whereas the $C_4$-symmetry profile consists of alternating sectors of width $\varphi = 45^{\circ}$. Likewise to the radial perturbation, we validate the RSE solution with the numerical one, obtained by COMSOL Multiphysics, as shown in Fig.~\ref{Draft_Figure 5}. Similarly to the previous cases, we present the reference solution for the unperturbed system.

In sectoral perturbation, different azimuthal harmonics are coupled. This coupling prevents a frequency shift or changes in the pressure field distribution for any single mode, unlike previous perturbation types. Consequently, the pressure field exhibits a complex form for both sectoral perturbation symmetries, as shown in Fig.~\ref{Draft_Figure 5}. One may observe that the asymmetric field distribution of the chosen mode is due to the difference between material parameters in the half-cylinder perturbation, which leads to a higher degree of field localization and the appearance of another field maximum for the half-cylinder profile in Fig.~\ref{Draft_Figure 5}(a). Similarly, the $ C_4$-symmetric mode chosen in Fig.~\ref{Draft_Figure 5}(c), is generally localized within the perturbed sector areas, but in contrast has a symmetric spatial distribution.

More importantly, the sectoral perturbation produces a significant shift in the eigenfrequency and a pronounced reduction in the quality factor $Q$ of the ``whispering-gallery'' modes. Increasing the number of perturbation sectors leads to a greater loss increase, as shown in the comparison in Figs.~\ref{Draft_Figure 5}(b) and \ref{Draft_Figure 5}(d). In scattering terms, the reduction of the Q factor originates from enhanced radiative leakage induced by the sectoral perturbation, which breaks the original symmetry and couples the resonant state to additional radiative channels. Since the perturbations $\Delta\rho$ and $\Delta\beta$ are real, no additional absorptive material losses are introduced. 

Here, we use a number of basis sizes equal to $N = 5694$ for the RSE calculation. One may observe RSE's high accuracy in comparison with numerical simulations obtained by COMSOL Multiphysics. Previously, we discussed the significance of the presence of cut poles for the RSE in the case of homogeneous and radial perturbations. However, low-$Q$ modes in sectoral perturbations do not require cut-mode contributions due to the high magnitude of the imaginary part of the eigenfrequency, as shown in panel (d) in Fig.~\ref{Draft_Figure 5}. That fact leads to a substantial reduction in numerical simulation time. Notably, the RSE system also influences the number of found eigenfrequencies in a finite time, unlike numerical simulation solutions, which give fewer numbers. This makes it possible to accurately calculate natural frequencies for resonators of arbitrary shape without solving the problem of eigenvalues for them.

\section{Conclusion}
\label{sec:conclusion}

We developed the complete RSE for open acoustic systems. This was achieved by formulating the GF of a two-dimensional cylindrical resonator, deriving analytic normalization for acoustic RSs, and obtaining the secular equations for their complex eigenfrequencies. Together, these results establish a rigorous perturbative framework. It parallels the optical RSE while adapting to the structure of acoustic fields and material parameters. The developed method accurately describes the influence of uniform, radial, and sectoral perturbations of density and compressibility. This enables controlled analysis of resonance shifts, mode mixing, and symmetry-induced coupling.

By expressing the perturbed eigenmodes in terms of the RSs of the unperturbed system, we obtain a unified framework for describing resonance shifts, mode hybridization, and symmetry-induced coupling. The excellent agreement with full-wave numerical simulations demonstrates the accuracy and robustness of the acoustic RSE for the considered classes of perturbations. These results establish the present formalism as a powerful analytical and computational tool for studying open acoustic resonators, slow-sound media, and structures with broken symmetries. Although demonstrated here for a two-dimensional cylindrical geometry, the underlying RSE formulation is not restricted to this specific system and can, in principle, be generalized to three-dimensional open acoustic resonators. Its perturbative structure also enables a natural integration with numerical eigenmode solvers, in which resonant states obtained for geometrically complex reference structures can be used as the unperturbed basis. Such a hybrid analytical-numerical implementation may extend the applicability of the acoustic RSE to complex resonator geometries, metamaterial design, and the analysis of non-Hermitian acoustic systems.

Here, we have focused on the eigenvalue problem, including the calculation of the complex eigenfrequencies and eigenvectors of open acoustic resonators. In Part~II of this series~\cite{domoratskii2026resonant-II}, we apply the developed RSE formalism to the scattering problem and use the same resonant-state basis to describe the acoustic scattering response.

\begin{acknowledgments}
The authors acknowledge financial support from the Russian Science Foundation (25-79-31027).
\end{acknowledgments}

\bibliography{references_PRB_complete}

\end{document}